\documentclass{article}
\usepackage{spconf, amsmath, epsfig, amssymb, booktabs, soul, color, xcolor}

\usepackage{hyperref}
\hypersetup{
    colorlinks=true,
    linkcolor=blue,
    citecolor=blue}

\title{Boosting Spatial-Spectral Masked Auto-Encoder Through Mining Redundant Spectra \\ for HSI-SAR/LiDAR Classification}

\name{Junyan Lin$^1$, Xuepeng Jin$^1$, Feng Gao$^1$, Junyu Dong$^1$, Hui Yu$^2$
\thanks{
This work was supported in part by the National Key R\&D Program of China under Grant 2022ZD0117201, and in part by the Leverhulme Trust through project VP1‐2020‐044.}}
\address{
$^1$School of Computer Science and Technology, Ocean University of China, Qingdao, China.\\
$^2$School of Creative Technologies, University of Portsmouth, Portsmouth, U.K.}

\begin{document}
\maketitle

\begin{abstract}

Although recent masked image modeling (MIM)-based HSI-LiDAR/SAR classification methods have gradually recognized the importance of the spectral information, they have not adequately addressed the redundancy among different spectra, resulting in information leakage during the pretraining stage. This issue directly impairs the representation ability of the model. To tackle the problem, we propose a new strategy, named Mining Redundant Spectra (MRS). Unlike randomly masking spectral bands, MRS selectively masks them by similarity to increase the reconstruction difficulty. Specifically, a random spectral band is chosen during pretraining, and the selected and highly similar bands are masked. Experimental results demonstrate that employing the MRS strategy during the pretraining stage effectively improves the accuracy of existing MIM-based methods on the Berlin and Houston 2018 datasets.

\end{abstract}


\section{Introduction}

With the rapid development of satellite sensors, multi-source remote sensing images have become increasingly accessible \cite{zhao23tgrs}. This undoubtedly enhances the feasibility of classifying multi-source remote sensing images, providing more reliable and accurate support for environmental monitoring, resource management, urban planning, and other fields \cite{wang20tgrs}.

\begin{figure}[t]
\centering
\includegraphics [width=3.3in]{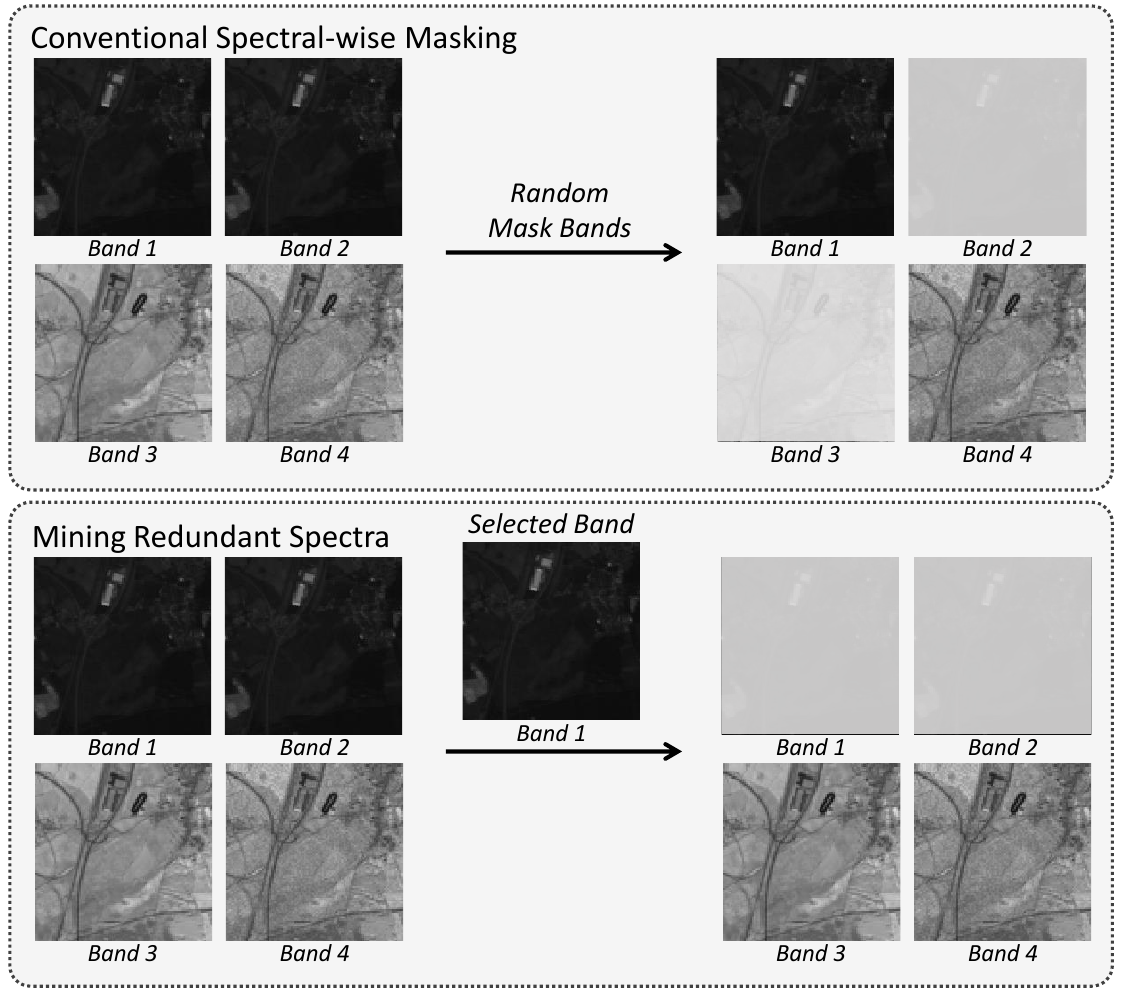}
\caption{Our proposed Mining Redundant Spectra (MRS) strategy.}
\label{paradigm} 
\end{figure}

In recent years, a large number of outstanding models have been developed using deep learning in the field of hyperspectral image (HSI) classification. However, the single-source classification may encounter challenges in certain scenarios. For instance, accurately identifying different classes composed of the same material using only hyperspectral images is challenging due to their similar spectral responses. Recently, many researchers have attempted to incorporate light detection and ranging (LiDAR) images, synthetic aperture radar (SAR) images, and other remote sensing images. This has effectively assisted models in distinguishing different classes composed of the same material and has greatly mitigated interference from external factors such as adverse environmental conditions. Gao et al. \cite{gaoyh22tgrs} developed a depthwise cross-attention module to capture both self-correlation and cross-correlation across various multisource data. Wang et al. \cite{9698196} introduced AM$^3$Net, incorporating an involution operator, as well as spectral-spatial mutual-guided modules. Li et al. \cite{c3} proposed a graph-based feature selective assignment network for multisource image feature fusion. Most of these methods are supervised models requiring labeled training data, posing a high labor demand. Masked Image Modeling (MIM) is an unsupervised representation learning strategy that learns universal feature representations through mask-reconstruction without labels. A representative MIM-based method in the field of multi-source remote sensing image classification is SS-MAE \cite{lin2023ss}, which employs spatial-wise masking and spectral-wise masking, respectively.


Although the spectral-wise masking strategy from SS-MAE considers spectral information, it still has shortcomings in overcoming information leakage caused by redundant spectral information. Fig. \ref{paradigm} illustrates four bands of a hyperspectral image, where band 1 and band 2, and band 3 and band 4 are highly similar. The spectral-wise masking from SS-MAE adopts a random masking strategy, making it easy for the model to reconstruct the masked bands based on highly similar bands during the reconstruction process when similar bands are not simultaneously masked. This leads to information leakage. To tackle this issue, we introduce the Mining Redundant Spectra (MRS) strategy, selectively masking spectral bands. More specifically, we randomly select a band, calculate the similarity of each band to it, and then mask both the selected band and those with high similarity. We boost the classification performance of SS-MAE on the Berlin dataset and Houston 2018 dataset, by simply replacing the masking strategy with MRS, which demonstrates the effectiveness of our MRS strategy.

\begin{figure}[t]
\centering
\includegraphics [width=3.3in]{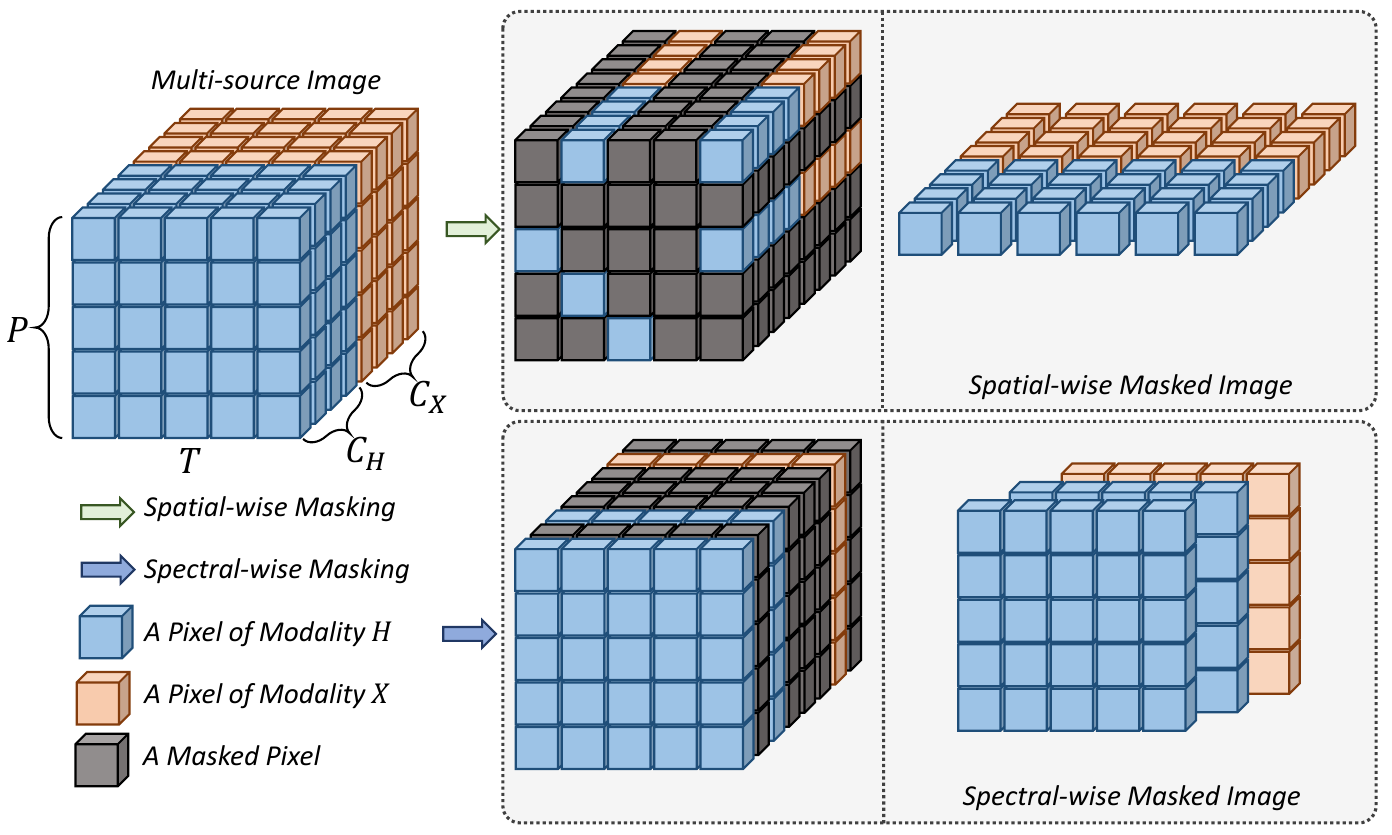}
\caption{Two different masking strategies in SS-MAE.}
\label{mask_strategy} 
\end{figure}

\section{Methodology}

In this section, we first present the overall process of masked image modeling in Section 2.1, encompassing the masking strategy and the reconstruction target. Subsequently, implementation details of the MRS strategy are provided in Section 2.2.

\subsection{Masked Image modeling}


The process of MIM-based methods for multi-source remote sensing image classification is usually as follows (Take the spectral-wise masking as an example): In the pretraining stage, for the input modalities $H \in \mathbb{R}^{C_H \times P \times P}$ and $X \in \mathbb{R}^{C_X \times P \times P}$, where $P$ represents the patch size, $C_H$ is the number of bands in modality $H$, and $C_X$ is the number of bands in modality $X$, a custom masking strategy (spectral-wise masking in here) is initially applied in the concatenation $T \in \mathbb{R}^{(C_H+C_X) \times P \times P}$ of two modalities to obtain the masked modalities $T_m \in \mathbb{R}^{(C_H^M+C_X^M) \times P \times P}$. Here, $C_H^M$ is the number of unmasked bands in $H$, and $C_X^M$ is the number of unmasked bands in $X$. Subsequently, an encoder maps $T_m$ to features $T_f \in \mathbb{R}^{(C_H^M+C_X^M) \times P \times P}$. Then, a decoder combines features $T_f$ and mask tokens $D \in \mathbb{R}^{(C_H+C_X-C_H^M-C_X^M) \times P \times P}$ to generate predicted outputs $T' \in \mathbb{R}^{(C_H+C_X) \times P \times P}$. Finally, the model is pretrained by calculating the reconstruction loss between the predicted outputs $T'$ and the concatenation $T$ of two modalities. In the training stage, the pre-trained encoder serves as the backbone network. The concatenation $T$ of two modalities is fed into the encoder to obtain features. Subsequently, a classifier is applied to generate classification results, and the model is fine-tuned using ground truth data.


\textbf{Masking Strategy:}
The key aspect of employing masked image modeling for learning universal representations is to mask input images, disrupting the image information, and then reconstructing the perturbed information through an autoencoder. Therefore, a proper information disruption strategy (i.e., masking strategy) is crucial. Due to differences in image characteristics, the computer vision field MIM-based methods do not emphasize spectral-wise information disruption and mostly use spatial-wise masking. However, in remote sensing, especially in hyperspectral images where a significant amount of information is embedded in the spectral dimension, relying solely on spatial dimension masking is inappropriate. To address this issue, SS-MAE \cite{lin2023ss}, a representative MIM-based method in the field of multi-source remote sensing image classification, introduces spectral-wise masking in addition to spatial-wise masking, which is illustrated in Fig. \ref{mask_strategy}. Specifically, for spectral-wise masking, assuming an input $T \in \mathbb{R}^{C_T \times P \times P}$, spectral-wise masked image $T_m^{spe} \in \mathbb{R}^{C_T^M \times P \times P}$ are obtained by randomly masking along the spectral dimension, where $C_T^M$ is the number of unmasked bands in $T_m^{spe}$. Through spectral-wise masking, SS-MAE achieves preliminary spectral representation learning.

\begin{figure*}[htb]
\centering
\includegraphics [width=0.95\textwidth]{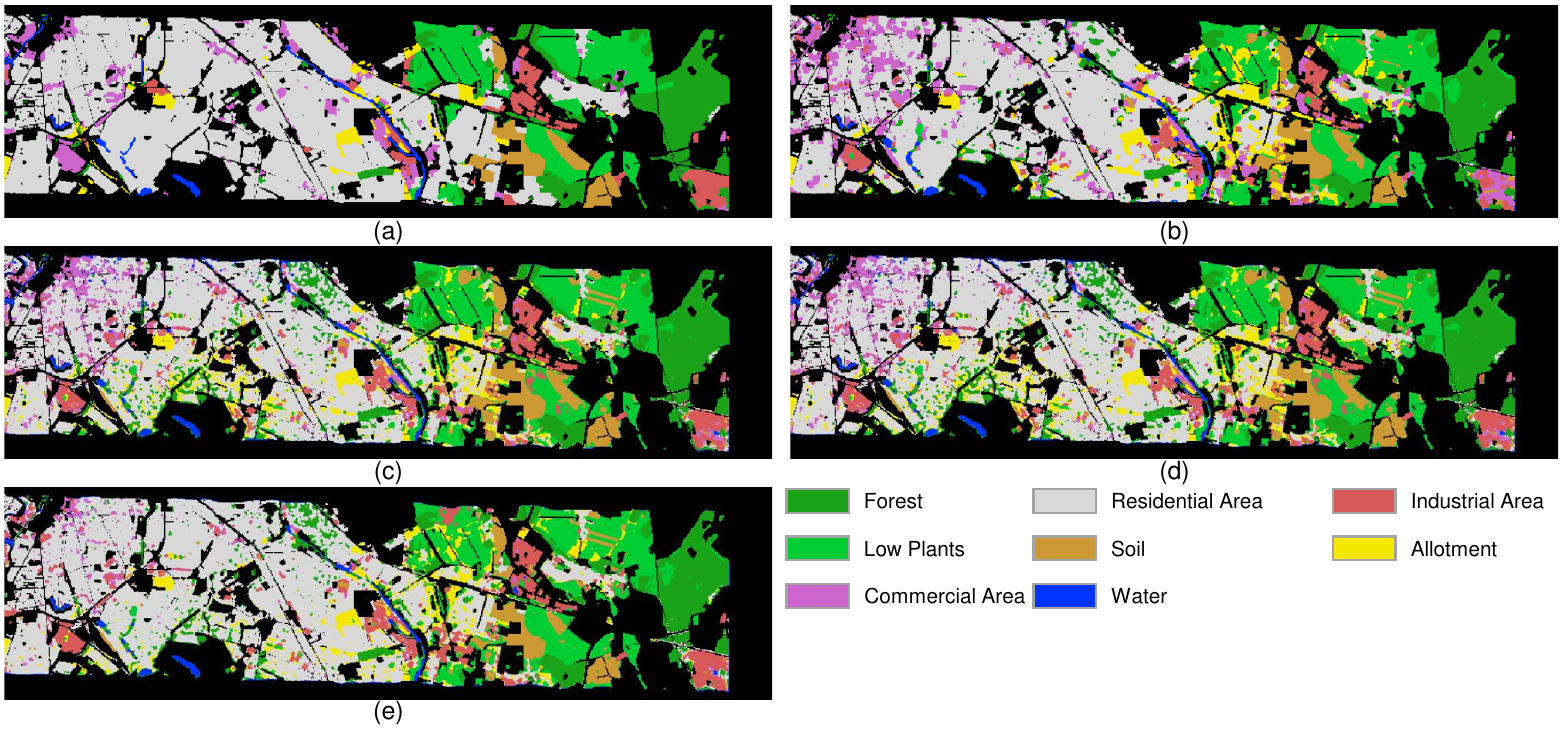}
\caption{Classification results of different methods for the Berlin dataset. (a) Ground Truth. (b) ExViT. (c) SS-MAE. (d) $\text{SS-MAE}^*$. (e) $\text{SS-MAE}'$.}
\label{berlin_fig_com}
\end{figure*}

\textbf{Reconstruction target:}
MIM-based methods reconstruct the input $T$ by predicting the pixel values of each masked region. The loss function is implemented by calculating the mean squared error (MSE) between the reconstructed image $T'$ and the input image $T$. We only compute the loss function within the masked regions.

\subsection{Mining Redundant Spectra}
Although SS-MAE has optimized the masking strategy along the spectral dimension, as illustrated in Fig. \ref{paradigm}, there still exists an issue of information leakage due to the similarity among some spectral bands. This issue adversely affects the spectral representation learning capability of the model. To address this concern, we propose a simple but effective strategy, named Mining Redundant Spectra (MRS). During the pre-training stage, MRS masks spectrally similar bands to avoid situations of referencing similar bands to reconstruct the masked band.

Specifically, we randomly selects a spectral band $ t \in \mathbb{R}^{1 \times P \times P} $ from the input $T \in \mathbb{R}^{C_T \times P \times P}$. Subsequently, we compute the cosine similarity $S \in \mathbb{R}^{C_T}$ between the selected spectral band $t$ and each of the $C_T$ spectral bands in $T$ as follow:
\begin{equation}
S_i = \frac{t \cdot T_i}{\|t\| \cdot \|T_i\|}, \quad \text{for } i = 1, 2, ..., C_T
\end{equation}
where $T_i$ represents the $i$-th spectral band in $T$.
Then, we utilize the computed cosine similarity $S$ to mask $t$ and selects the top $R(\%)$ most similar spectral bands, where $R$ is the mask ratio. It can be formulated as:
\begin{equation}
    T_m^{mrs} = \text{MaskTop}(R, S, T)
\end{equation}
where $T_m^{mrs} \in \mathbb{R}^{(C_T \times (1-R)) \times P \times P}$ is the masked image, $\text{Top}(R, S, T)$ represents a function that masks the top $R$ similar bands in $S$ by the similarity scores in vector $S$.

When the mask ratio $R$ is the same for spectral-wise masking and Mining Redundant Spectra (MRS), the dimensions of $T_{\text{m\_spe}}$ and $T_{\text{m\_mrs}}$ will also be identical. Therefore, we can readily substitute the spectral-wise masking in SS-MAE with our MRS to prevent information leakage.

\begin{table}[]
\centering
\caption{Classification Performance of Different Models on the Berlin Dataset.}
\vspace{3mm}
\small
\begin{tabular}{c|c c c c}
\toprule
Class & ExViT & $\mathcal{M}$ & $\mathcal{M}^*$ & $\mathcal{M}'$ \\
\midrule
	Forest & 78.01 & 79.53 & 79.35 & 78.21 \\ 
	Residential area &74.05 & 70.99 & 76.04 & 79.70 \\ 
        Industrial area & 39.48 & 63.06 & 68.51 & 67.90 \\
        Low plants & 84.15 & 80.19 & 79.17 & 79.31 \\
        Soil & 88.03 & 92.87 & 91.59 & 76.75 \\
        Allotment & 70.00 & 64.60 & 56.86 & 62.17 \\
        Commercial area & 38.18 & 26.37 & 19.75 & 11.47 \\
        Water & 56.41 & 68.40 & 66.87 & 61.78 \\
\midrule
OA & 72.63 & 71.15 & 73.43 & 74.51 \\ 
\bottomrule
\end{tabular}
\label{table_berlin}
\end{table}

\section{Experimental Results and Analysis}

To verify the effectiveness of our proposed masking strategy MRS, extensive experiments are carried out on the Berlin dataset and Houston 2018 dataset. Berlin dataset is used for hyperspectral and SAR data classification, encompassing both urban and rural areas in Berlin. The hyperspectral imagery (HSI) comprises 244 spectral bands within the wavelength range of 400–2500 nm. Houston 2018 dataset is used for hyperspectral and LiDAR data classification, covering the University of Houston campus and its neighboring urban surroundings. The HSI contains 48 bands in the wavelength range of 380–1050 nm.

\begin{table}[ht]
\centering
\caption{Classification performance of different models on the Houston 2018 dataset.}
\vspace{3mm}
\small
\begin{tabular}{c|c c c c}
\toprule
Class & ExViT & $\mathcal{M}$ & $\mathcal{M}^*$ & $\mathcal{M}'$ \\
\midrule
    Health grass &  93.65 &  90.53 & 94.00&97.90 \\      
    Stressed grass & 95.44 &  95.56& 95.34&92.81\\ 
    Artificial turf  & 100.00 & 100.00& 100.00&100.00\\
    Evergreen trees & 98.52 &  98.23 & 98.55&98.92\\ 
    Deciduous trees &99.25 & 96.62& 98.15&98.46\\ 
    Bare earth &99.92 &  100.00& 99.84&99.90\\ 
    Water & 100.00 &  100.00& 100.00&100.00\\ 
    Res. buildings  &96.88 &  95.89& 94.98&95.15\\ 
    Non-res. buildings & 94.87 &  95.85& 96.27&95.99\\ 
    Roads & 80.51 & 78.77& 82.97&79.34\\ 
    Sidewalks & 80.27 & 78.29& 74.10&83.65\\ 
    Crosswalks & 96.76 & 97.78& 89.57&94.92\\ 
    Major thoroughfares & 82.11 & 80.20& 83.21&85.73\\ 
    Highways & 84.10 &98.28& 98.08&98.92\\ 
    Railways & 99.81 &100.00& 99.73&99.90\\ 
    Paved parking lots & 99.08 & 95.36& 96.36&97.34\\ 
    Unpaved parking lots & 100.00 &  100.00& 100.00&100.00\\ 
    Cars & 97.56 &98.19 & 97.20&99.14\\ 
    Trains & 99.90 &99.46& 100.00&100.00\\ 
    Stadium seats & 99.94 & 100.00& 100.00&99.97\\
\midrule
OA & 91.87 & 91.91& 92.43 & 92.87 \\ 
\bottomrule
\end{tabular}
\label{table_houston}
\end{table}

In this paper, we only replace the spectral-wise masking of SS-MAE with MRS for pre-training. After pre-training, we fine-tuned SS-MAE to obtain $\text{SS-MAE}'$. We compare the classification performance of $\text{SS-MAE}'$, $\text{SS-MAE}^*$ (pre-trained using the original masking strategy), SS-MAE (without pre-training), and the state-of-the-art method, ExViT \cite{yao2023extended}. The Overall Accuracy (OA) values for all methods are presented in Table \ref{table_berlin} and Table \ref{table_houston},  where $\mathcal{M}$, $\mathcal{M}^*$, and $\mathcal{M}'$ respectively represent $\text{SS-MAE}$, $\text{SS-MAE}^*$, and $\text{SS-MAE}'$. It is evident from the results that $\text{SS-MAE}'$ pre-trained with our MRS outperforms other methods. We also visualize the classification results of the Berlin dataset, as shown in Fig. \ref{berlin_fig_com}. Fig. \ref{berlin_fig_com}(e) is the classification results of $\text{SS-MAE}'$, which closely align with the ground truth. Both quantitative and qualitative experimental results confirm the effectiveness of our proposed MRS strategy and its convenience for plug-and-play applications.

\section{Conclusion}
In this paper, we propose a strategy called Mining Redundant Spectra (MRS) to alleviate the issue of information leakage in the process of spectral-wise mask-reconstruction. Specifically, we discard the strategy of spectral-wise random masking. In each iteration, we select a band as a base, filter out bands similar to it, and after masking these bands, we can reconstruct the masked bands without hints from similar bands. By simply replacing the masking strategy in SS-MAE with our MRS, we achieve significant improvements in classification performance on the Berlin dataset and Houston 2018 dataset. This demonstrates the effectiveness of our MRS strategy and its convenience for plug-and-play applications in MIM-based methods for images with multi-bands.

\end{document}